\documentclass[3p,times,twocolumn]{elsarticle}

\usepackage{ecrc}
\usepackage{color}
\usepackage{xspace}
\usepackage{amsmath}
\usepackage{ifthen} 
\newboolean{uprightparticles}
\setboolean{uprightparticles}{false} 

\def\ux85 {\mbox{UX85}\xspace}

\ifthenelse{\boolean{uprightparticles}}%
{

 \def\Pmu         {\ensuremath{\upmu}\xspace}

 \def\Ppi         {\ensuremath{\uppi}\xspace}

 \def\PDelta      {\ensuremath{\Delta}\xspace}                 
 \def\PXi      {\ensuremath{\Xi}\xspace}                 
 \def\PLambda      {\ensuremath{\Lambda}\xspace}                 
 \def\PSigma      {\ensuremath{\Sigma}\xspace}                 
 \def\POmega      {\ensuremath{\Omega}\xspace}                 
 \def\PUpsilon      {\ensuremath{\Upsilon}\xspace}                 
 
 \def\PB      {\ensuremath{\mathrm{B}}\xspace}                 
                  
 \def\PD      {\ensuremath{\mathrm{D}}\xspace}

 \def\PK      {\ensuremath{\mathrm{K}}\xspace}

 \def\Pi      {\ensuremath{\mathrm{i}}\xspace}

 \def\Ps      {\ensuremath{\mathrm{s}}\xspace}

}
{

 \def\Pmu         {\ensuremath{\mu}\xspace}

 \def\Ppi         {\ensuremath{\pi}\xspace}

 \mathchardef\PDelta="7101
 \mathchardef\PXi="7104
 \mathchardef\PLambda="7103
 \mathchardef\PSigma="7106
 \mathchardef\POmega="710A
 \mathchardef\PUpsilon="7107
                  
 \def\PB      {\ensuremath{B}\xspace}                 
                  
 \def\PD      {\ensuremath{D}\xspace}

 \def\PK      {\ensuremath{K}\xspace}

 \def\Pi      {\ensuremath{i}\xspace}

 \def\Ps      {\ensuremath{s}\xspace}

}

\def\mup        {\ensuremath{\Pmu^+}\xspace}
\def\mun        {\ensuremath{\Pmu^-}\xspace} 

\def\ellell     {\ensuremath{\ell^+ \ell^-}\xspace}

\def\squark    {\ensuremath{\Ps}\xspace}

\def\pion  {\ensuremath{\Ppi}\xspace}

\def\pim   {\ensuremath{\pion^-}\xspace}

\def\kaon  {\ensuremath{\PK}\xspace}
  \def\Kbar  {\kern 0.2em\overline{\kern -0.2em \PK}{}\xspace}

\def\Kz    {\ensuremath{\kaon^0}\xspace}
\def\Kzb   {\ensuremath{\Kbar^0}\xspace}
\def\KzKzb {\ensuremath{\Kz \kern -0.16em \Kzb}\xspace}
\def\Kp    {\ensuremath{\kaon^+}\xspace}
\def\Km    {\ensuremath{\kaon^-}\xspace}

\def\KpKm  {\ensuremath{\Kp \kern -0.16em \Km}\xspace}

\def\Kstarz  {\ensuremath{\kaon^{*0}}\xspace}

  \def\Dbar    {\kern 0.2em\overline{\kern -0.2em \PD}{}\xspace}
\def\D       {\ensuremath{\PD}\xspace}

\def\Dz      {\ensuremath{\D^0}\xspace}
\def\Dzb     {\ensuremath{\Dbar^0}\xspace}
\def\DzDzb   {\ensuremath{\Dz {\kern -0.16em \Dzb}}\xspace}
\def\Dp      {\ensuremath{\D^+}\xspace}
\def\Dm      {\ensuremath{\D^-}\xspace}

\def\DpDm    {\ensuremath{\Dp {\kern -0.16em \Dm}}\xspace}

\def\Ds      {\ensuremath{\D^-_\squark}\xspace}

\def\B       {\ensuremath{\PB}\xspace}
  \def\Bbar    {\kern 0.18em\overline{\kern -0.18em \PB}{}\xspace}

\def\Bu      {\ensuremath{\B^+}\xspace}

\def\Bp      {\ensuremath{\Bu}\xspace}

\def\Bd      {\ensuremath{\B^0}\xspace}
\def\Bs      {\ensuremath{\B^0_\squark}\xspace}

  \def\Y#1S{\ensuremath{\PUpsilon{(#1S)}}\xspace}

\def\Lbar {\ensuremath{\kern 0.1em\overline{\kern -0.1em\PLambda}}\xspace}

\def\BF         {{\ensuremath{\cal B}\xspace}}

\newcommand{\decay}[2]{\ensuremath{#1\!\to #2}\xspace}         
\def\ra                 {\ensuremath{\rightarrow}\xspace}
\def\to                 {\ensuremath{\rightarrow}\xspace}

\def\BdToKstmm    {\decay{\Bd}{\Kstarz\mup\mun}}

\def\AT#1     {\ensuremath{A_{\mathrm{T}}^{#1}}\xspace}           

\def\Bsmm     {\decay{\Bs}{\mup\mun}}
\def\Bdmm     {\decay{\Bd}{\mup\mun}}

\def\C#1      {\ensuremath{\mathcal{C}_{#1}}\xspace}                       
\def\Cp#1     {\ensuremath{\mathcal{C}_{#1}^{'}}\xspace}                    
\def\Ceff#1   {\ensuremath{\mathcal{C}_{#1}^{\mathrm{(eff)}}}\xspace}        
\def\Cpeff#1  {\ensuremath{\mathcal{C}_{#1}^{'\mathrm{(eff)}}}\xspace}       
\def\Ope#1    {\ensuremath{\mathcal{O}_{#1}}\xspace}                       
\def\Opep#1   {\ensuremath{\mathcal{O}_{#1}^{'}}\xspace}                    

\newcommand{\tev}{\ensuremath{\mathrm{\,Te\kern -0.1em V}}\xspace}
\newcommand{\gev}{\ensuremath{\mathrm{\,Ge\kern -0.1em V}}\xspace}
\newcommand{\mev}{\ensuremath{\mathrm{\,Me\kern -0.1em V}}\xspace}
\newcommand{\kev}{\ensuremath{\mathrm{\,ke\kern -0.1em V}}\xspace}
\newcommand{\ev}{\ensuremath{\mathrm{\,e\kern -0.1em V}}\xspace}
\newcommand{\gevc}{\ensuremath{{\mathrm{\,Ge\kern -0.1em V\!/}c}}\xspace}
\newcommand{\mevc}{\ensuremath{{\mathrm{\,Me\kern -0.1em V\!/}c}}\xspace}
\newcommand{\gevcc}{\ensuremath{{\mathrm{\,Ge\kern -0.1em V\!/}c^2}}\xspace}
\newcommand{\gevgevcccc}{\ensuremath{{\mathrm{\,Ge\kern -0.1em V^2\!/}c^4}}\xspace}
\newcommand{\mevcc}{\ensuremath{{\mathrm{\,Me\kern -0.1em V\!/}c^2}}\xspace}

\def\mub{\ensuremath{\rm \,\Pmu b}\xspace}

\def\invfb   {\ensuremath{\mbox{\,fb}^{-1}}\xspace}

\def\gsim{{~\raise.15em\hbox{$>$}\kern-.85em
          \lower.35em\hbox{$\sim$}~}\xspace}
\def\lsim{{~\raise.15em\hbox{$<$}\kern-.85em
          \lower.35em\hbox{$\sim$}~}\xspace}

\def\pt         {\mbox{$p_{\rm T}$}\xspace}

\def\tell1  {TELL1\xspace}
\def\ukl1   {UKL1\xspace}

\newcommand{\Bhh}{\ensuremath{B^0_{(s)}\to h^+h'^-}\xspace}

\newcommand{\Bsdmm}{\ensuremath{\ensuremath{B^0_{s,d}}\to\mu^+\mu^-}\xspace}

\newcommand{\tmmm}{\ensuremath{\tau^-\to \mu^+\mu^-\mu^-}\xspace}

\newcommand{\BRof}[1]{\ensuremath{{\cal B}(#1)}\xspace}

\volume{00}

\firstpage{1}

\journalname{Nuclear Physics B Proceedings Supplement}

\runauth{J. Albrecht}

\jid{nuphbp}

\jnltitlelogo{Nuclear Physics B Proceedings Supplement}

\usepackage{amssymb}

\usepackage{lineno}

\usepackage[figuresright]{rotating}

\begin{document}

\begin{frontmatter}



\dochead{}

\title{Search for New Physics in rare decays at LHCb}

\author{Johannes Albrecht\fnref{label2}}
\fntext[label2]{On behalf of the LHCb collaboration}
\address{Johannes.Albrecht@cern.ch\\CERN, Geneva, Switzerland}




\begin{abstract}

Rare heavy flavor decays provide stringent tests of the Standard Model
of particle physics and allow to test for possible new Physics
scenarios.
The LHCb experiment at CERN is the ideal place for these searches as
it has recorded the worlds largest sample of
beauty mesons. The status of the rare decay analyses with 1\invfb of
$\sqrt s = 7\tev $ of $pp$--collisions collected by the LHCb experiment
in 2011 is reviewed.
The worlds most precise measurements of the angular structure of
\BdToKstmm decays is discussed, as well as the isospin asymmetry
measurement in $\decay{B}{\kaon^{(*)} \mup\mun}$ decays. The most
stringent upper exclusion limit on the branching fraction of \Bsmm decays is shown, as well as
searches for lepton number and lepton flavor violating 
processes. 

\end{abstract}


\end{frontmatter}


\section{Introduction}
\label{sec:intro}

Rare processes which proceed via flavor changing neutral currents (FCNC) 
are forbidden at tree level in the Standard Model (SM). They can
proceed via loop level electroweak ($Z^0, \gamma$) penguin or box diagrams. In
extensions to the SM, new virtual particles can enter at loop level,
modifying the amplitude of the process or the Lorentz structure of the
decay vertex. Possible deviations from the SM predictions on
  branching fractions or angular distributions could lead to
the discovery of physics beyond the SM.

This article reviews some of the most sensitive probes for possible
extensions of the Standard Model that were measured by the LHCb
collaboration with a dataset of 1\invfb of $\sqrt s = 7\tev$ of
$pp$--collisions collected in 2011. 
An angular analysis of \BdToKstmm decays\footnote{In
   this proceedings, the inclusion of charge conjugate states are
   implicit.} allows a stringent test of
the Lorentz structure of the electroweak penguin process and is
particularly sensitivity to right-handed currents.   
The measurement of the decay rate of \Bsdmm decays are highly sensitive
to scalar and pseudoscalar currents, which are non-existent in the
SM. The article closes with a discussion of searches for lepton
number and lepton flavor violation and $B$-hadron and
$\tau^\pm$-lepton decays. 

The rare decay processes presented here provide a complementary approach to direct
searches at the general purpose detectors and can give sensitivity
to new particles at higher mass scales than those accessible directly.

\section{Electroweak penguin decays}
\label{ewp}

\subsection{Angular analysis of \BdToKstmm decays}

The decay \BdToKstmm allows the construction of several observables with
small hadronic uncertainties, that are sensitive to physics beyond the
Standard Model (see \cite{Ali1991505,Altmannshofer:2008dz} and references therein). These
observables include~\cite{Altmannshofer:2008dz,Kruger:2005ep,Bobeth:2008ij,Bobeth:2010wg}
\begin{figure}[!pb]
\centering
\includegraphics[width=0.47\textwidth]{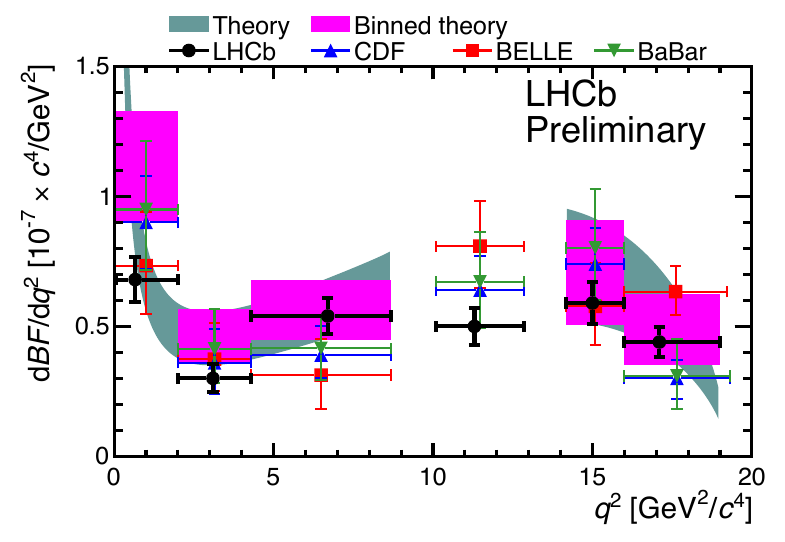}
\caption
{Differential decay rate of \BdToKstmm, measured by the BABAR~\cite{n25},
  Belle~\cite{Wei:2009zv}, CDF~\cite{n28} and LHCb~\cite{LHCb-CONF-2012-008}
  experiments. The SM prediction, from Ref.~\cite{Bobeth:2010wg}, is also
  shown. Figure reproduced from Ref.~\cite{LHCb-CONF-2012-008}.} 
\label{fig:kst_q2}
\end{figure} 
\begin{itemize}
\item \textbf{$A_{FB}$}, the forward-backward asymmetry of the
  dimuon system,
\item \textbf{$F_L$}, the fraction of \Kstarz longitudinal polarization,
\item \textbf{$S_3$}, the transverse asymmetry, which is also often
  referred to as $\frac{1}{2}(1-F_L)A_T^2$ and 
\item \textbf{$S_9$}, a $CP$ averaged quantity corresponding to the
  imaginary component of the product of the longitudinal and
  transverse amplitudes of the \Kstarz.
\end{itemize}
 
The LHCb collaboration performs an angular analysis in bins of 
the squared dimuon invariant mass  
($q^2$) and the three angles $\theta_l$, $\theta_k$ and
$\phi$. $\theta_l$ is defined as the angle between the \mup and the
\Bd in the dimuon rest frame, $\theta_k$ as angle between the kaon and
the \Bd in \Kstarz rest frame and $\phi$ as angle between the plane
spanned by the dimuon system and the \Kstarz decay plane.

\begin{figure*}[!htb]
\centering
\includegraphics[width=0.94\textwidth]{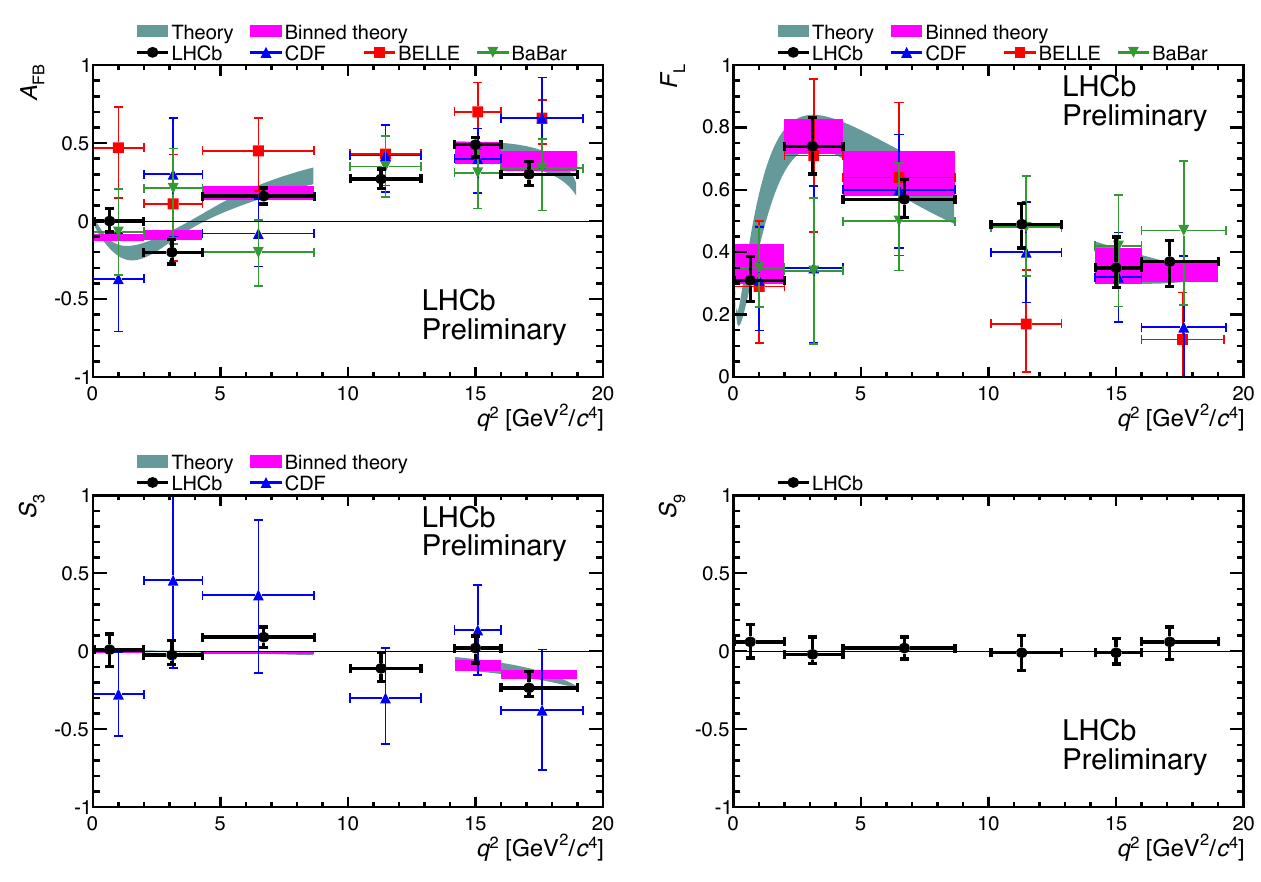}
\caption
{The observables $A_{FB}$, $F_L$, $S_3$ and $S_9$ measured in
  \BdToKstmm decays by the BABAR~\cite{n25}, 
  Belle~\cite{Wei:2009zv}, CDF~\cite{n28} and LHCb~\cite{LHCb-CONF-2012-008}
  experiments. The SM prediction, from Ref.~\cite{Bobeth:2010wg}, is also
  shown. Figure reproduced from Ref.~\cite{LHCb-CONF-2012-008}.}
\label{fig:kst_obs}
\end{figure*}

The differential branching ratio as a function of $q^2$ is shown in
Fig.~\ref{fig:kst_q2}, together with recent measurements from other
collaborations. The measurements from the   
BABAR, Belle, CDF and LHCb collaborations are consistent with each other.
Measurements of the observables $A_{FB}$, $F_L$, $S_3$ and $S_9$
 are shown in
Fig.~\ref{fig:kst_obs}. These are the most precise measurements of
these observables and no deviations from the SM predictions have been
seen. 

The zero-crossing point of $A_{FB}$, $q_0$, is a particularly sensitive probe
for NP and, as the form factor uncertainties cancel at first order, it
is theoretically very clean. The LHCb collaboration has
reported the worlds first measurement as
$q_0=4.9^{+1.1}_{-1.3}$\,GeV$^2$/c$^4$, in good agreement with the SM
prediction. This measurement strongly disfavours scenarios with a
flipped sign of the Wilson coefficient $C_7$.

\subsection{Isospin asymmetry in $\decay{B}{\kaon^{(*)} \mup\mun}$}

\begin{figure}[!htb]
\centering
\includegraphics[width=0.47\textwidth,height=0.3\textwidth]{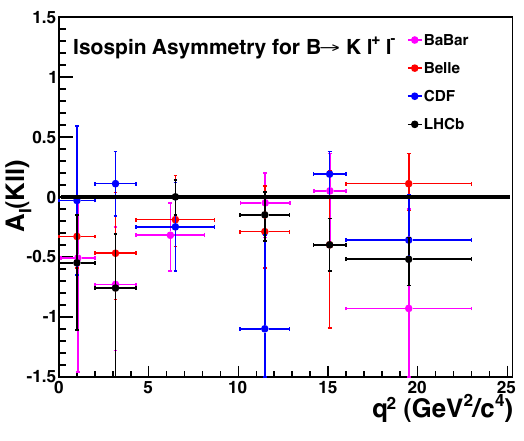}\\
\includegraphics[width=0.47\textwidth, height=0.3\textwidth]{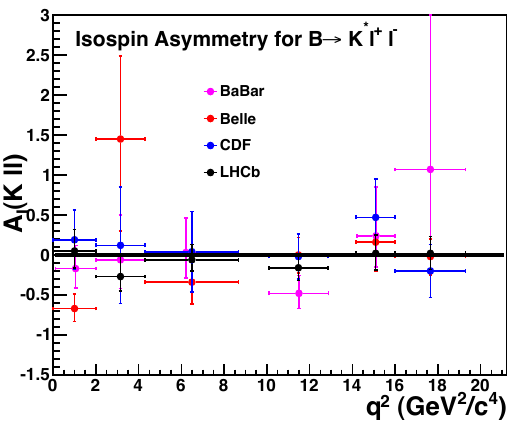}
\caption
{Isospin asymmetries for the decays $\decay{B}{\kaon^{(*)} \ellell}$
  measured in the muonic channels by CDF~\cite{n28} and LHCb~\cite{Aaij:2012cq}
  as well as in both the electron and the muon channel by
  BABAR~\cite{n25} and Belle~\cite{Wei:2009zv}. Figure reproduced from
  Ref.~\cite{Aaij:2012cq}.} 
\label{fig:iso}
\end{figure}

The isospin asymmetry of the decays $\decay{B}{\kaon^{(*)} \mup\mun}$,
$A_I$, is defined as
\begin{equation}
A_I=\frac
{\BRof{\decay{\Bd}{\kaon^{(*)0} \mup\mun}} - \frac{\tau_0}{\tau_+}
  \BRof{\decay{\Bu}{\kaon^{(*)+} \mup\mun}}}
{\BRof{\decay{\Bd}{\kaon^{(*)0} \mup\mun}} + \frac{\tau_0}{\tau_+}
  \BRof{\decay{\Bu}{\kaon^{(*)+} \mup\mun}}}\, ,
\end{equation}
where $\tau_{0,+}$ is the lifetime of the \Bd and \Bu meson
respectively. 
For the $\decay{B}{\kaon^{*} \mup\mun}$ system, in the SM, $A_I$ is
predicted to be $-0.01$~\cite{Feldmann:2002iw}
with 
a slight increase at low values of $q^2$. 
For the $\decay{B}{\kaon \mup\mun}$ system, no SM calculation of $A_I$
exists, but it is similarly expected to be close to zero. 
The most precise measurement of $A_I$ is performed by the LHCb
collaboration~\cite{Aaij:2012cq}, it is shown in Fig.~\ref{fig:iso} together with
previous measurements of this quantity. All measurements are
consistent with each other and the $\decay{B}{\kaon^{*} \mup\mun}$
measurement is also consistent  with the SM prediction. The
$\decay{B}{\kaon \mup\mun}$ measurements are consistent amongst the
experiments but are less consistent with the naive expectation that $A_I=0$.
The deviation from zero has a significance of greater than four
standard deviations~\cite{Aaij:2012cq}.

\section{Searches for very rare and forbidden decays}
\label{vrd}

\subsection{Search for an extended Higgs sector in \Bsmm}

 
Precise measurements of the branching fractions of the two FCNC decays
\Bsmm and \Bdmm belong to the most promising measurements for a possible
discovery of a theory beyond the SM. 
These decays are strongly suppressed by loop and helicity factors,
making the SM branching fraction small~\cite{Buras:2012ru}:
$\BRof{\Bsmm}= (3.23 \pm 0.27) \times 10^{-9}$ and  $\BRof{\Bdmm} =
(0.11 \pm 0.01) \times 10^{-9}$.  
Taking the finite width difference of the \Bs system into account, the
time integrated branching fraction is evaluated to be~\cite{deBruyn:2012wk}
\begin{equation}
\BRof{\Bsmm}_{SM, \langle t \rangle} = 3.4 \times 10^{-9}\, . \label{eq:smbsmm}
\end{equation}
Enhancements of the decay rates of these decays are predicted in a
variety of different New Physics models, a summary is given in
Ref.~\cite{Bediaga:2012py}. For example, in the minimal 
supersymmetric Standard Model (MSSM), the enhancement is proportional
to $\tan^6\beta$, where $\tan\beta$ is the ratio of the vacuum
expectation values of the two Higgs fields. For large values of $\tan
\beta$, this search belongs to the most sensitive probes for physics
beyond the SM which can be performed at collider experiments. A
review of the experimental status of the searches for \Bsdmm can be
found in~\cite{Albrecht:2012hp}.

\vskip 0.2cm

The measurements presented here use 1\invfb of data recorded by the
LHCb experiment in 2011. Assuming the SM
branching ratio, about 12 (1.3) \Bs (\Bd) decays  are expected to be
triggered, reconstructed and selected in the analyzed dataset.
 
\vskip 0.2cm

The first step of the analysis is a simple selection, which removes
the dominant part of the background and keeps about 60\% of the
reconstructed signal events. 
As second step, a preselection, based on a Boosted Decision Tree (BDT) reduces 80\% of the
remaining background while retaining 92\% of the signal.

Each event is then given a probability to be signal or background in a
two-dimensional probability space defined by the dimuon invariant mass
and a multivariate discriminant likelihood. This likelihood combines 
kinematic and topological variables of the $B^0_{(s)}$ decay using a
BDT. The BDT is defined and trained on
simulated events for both signal and background. The signal BDT shape
is then calibrated using decays of the type $B^0_{(s)} \rightarrow h^+
h^{'-}$, where $h^\pm$ represents a $K^\pm$ or $\pi^\pm$. These
decays have an identical topology to the signal. The invariant mass
resolution is calibrated with an interpolation of $J/\psi$, $\psi(2S)$
and $\Upsilon(1S)$, $\Upsilon(2S)$ and $\Upsilon(3S)$ decays to two muons. The background shapes
are calibrated simultaneously in the mass and the BDT using the
invariant mass sidebands. This procedure ensures that even though the BDT
is defined using simulated events, the result will not be biased by
discrepancies between data and simulation. 

\begin{figure}[!b]
\centering
\includegraphics[width=0.47\textwidth,height=0.35\textwidth]{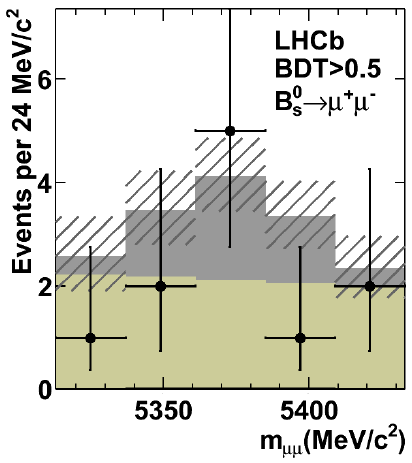}\\
\includegraphics[width=0.47\textwidth, height=0.35\textwidth]{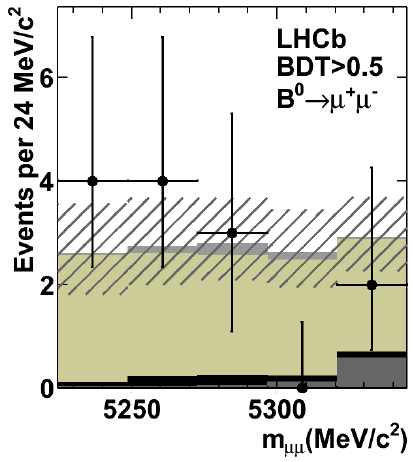}
\caption
{Distribution of selected signal candidates.   
  Events observed in LHCb in the \Bs channel (top) and the
  \Bd channel (bottom) for BDT$ > 0.5$ and expectation for, from top, SM signal
  (grey), combinatorial background (light grey), \Bhh background
  (black) and cross-feed between both modes (dark grey). The hatched
  area depicts the uncertainty on the total background expectation.
  Figure reproduced from Ref.~\cite{Aaij:2012ac}.} 
\label{fig:Bsmm_sig}
\end{figure}

The number of expected signal events is evaluated by normalizing with
channels of known branching fraction. Three independent channels are
used: $B^+\rightarrow J/\psi K^+$, $B^0_s\rightarrow J/\psi \phi$ and
$B^0 \rightarrow  K^+ \pi^-$. The first two decays have similar
trigger and muon identification efficiency to the signal but a
different number of particles in the final state, while the third
channel has the same two-body topology but is selected with a
hadronic trigger. The event selection for these
channels is specifically designed to be as close as possible to the
signal selection. The ratios of reconstruction and selection
efficiencies are estimated from the simulation, while the ratios of
trigger efficiencies on selected events are determined from data 
The observed pattern of events in the high BDT range is shown in
Fig.~\ref{fig:Bsmm_sig} for \Bsmm (top) and \Bdmm (bottom). A moderate
excess over the background expectations is seen in the \Bs
channel. This excess is consistent with the SM prediction. 
No excess is seem in the \Bd channel.

The compatibility of the observed distribution of events with a given
branching fraction hypothesis is computed using the CLs
method~\cite{Junk:1999kv,0954-3899-28-10-313}. 
The measured upper limit for the branching ratio is at 95\% confidence
level (CL)
\begin{eqnarray}
\BRof{\Bsmm}_{LHCb} &<& 4.5 \times 10^{-9}\, \textrm{ and }\\
\BRof{\Bdmm}_{LHCb} &<& 1.0 \times 10^{-9}\, ,
\end{eqnarray}
which are the worlds best upper exclusion limit on the branching
fraction of this decay.
A combination~\cite{LHCb-CONF-2012-017} of this measurement with the
ATLAS and CMS upper exclusion limits yields at 95\% CL 
\begin{eqnarray}
\BRof{\Bsmm}_{LHC} &<& 4.2 \times 10^{-9}\, \textrm{ and }\\
\BRof{\Bdmm}_{LHC} &<& 0.8 \times 10^{-9}\, ,
\end{eqnarray}
which is only a factor 20\% above the SM prediction given in
Eq.~\ref{eq:smbsmm}. This puts tight constraints on various
extensions of the Standard Model, especially on supersymmetric models
at high values of $\tan \beta$.

The CMS and LHCb collaborations have excellent prospects to observe
the decay \Bsmm with the dataset collected in 2012. This observation,
and the precision measurement of \BRof\Bsmm in the coming years
will allow to put strong constraints on the scalar sector of any extension of
the Standard Model. The next step will be to limit and then measure
the ratio of the decay rates of \Bsmm/\Bdmm, which allows a stringent
test of the hypothesis of minimal flavor violation and a good
discrimination between various extensions of the Standard Model.

In the absence of an observation, limits on \linebreak \BRof{\Bsdmm} are
complementary to those provided by high \pt experiments.
The interplay between both channels allows the SUSY parameter space to
be optimally constrained.

\subsection{Searches for an Majorana neutrinos in \B decays}

A search for Majorana neutrinos with a mass of $\mathcal{O}(1\gev)$ can be made
in lepton number violating (LNV) $B$ and $D$ meson decays. These indirect
searches are performed by analysing the decay rate of processes as
$\Bp \ra \pim
\mup\mup$~\cite{Shaposhnikov:2008pf,Atre:2009rg}.   
These same sign di-leptonic decays can only occur via exchange of heavy
Majorana neutrinos.  Resonant production may be possible if the heavy
neutrino is kinematically accessible, which could put the rates of
these decays within reach of the present or future LHCb luminosity. 
Alternatively, limits on these LNV processes, together with low energy
neutrino data, results in better constraints for neutrino masses
and mixing parameters in models with extended neutrino sectors. 

Using 0.4\invfb of integrated luminosity from LHCb, limits have been
set on the branching fraction of \linebreak $B^{+} \to
D_{(s)}^{-}\mup\mup$ decays at the level of a few times $10^{-7}$ and
on \decay{\Bp}{\pim\mup\mup} at the level of $1\times
10^{-8}$~\cite{LHCb-PAPER-2011-009, LHCb-PAPER-2011-038}. These
branching fraction limits imply a limit on the coupling
$|V_{\mu 4}|$ between $\nu_\mu$ and a Majorana neutrino with a mass in
the range $1 < m_{N} < 4 \gevcc$ of $|V_{\mu 4}|^{2} < 5\times
10^{-5}$.

\subsection{Search for lepton flavor violation in \tmmm}

Lepton flavor violating (LFV) $\tau^-$ decays are forbidden in the classical
SM and are vanishing small after the extension of the SM with neutrino
mixing. Many New Physics models predict enhancements, up to observable
values which are close to the current experimental bounds. 

The neutrinoless decay $\tau^-\to\mu^+\mu^-\mu^-$ is a particular
sensitive mode in which to search for LFV at LHCb as the experimental
signature with the three muon final state is very clean. 
The inclusive $\tau^-$ production cross-section at LHCb is very large,
about $80\mub$. The composition of the $\tau^-$ production can be
calculated from the $b\bar{b}$ and $c\bar{c}$~\cite{Aaij:2011jh} cross-sections measured
at the LHCb experiment and the inclusive branching ratios $b\ra \tau$
and $c\ra \tau$~\cite{pdg10}. About 80\% of the produced
$\tau^-$--leptons originate from $\Ds$ decays. 

LHCb has performed a search for the decay $\tau^-\to\mu^+\mu^-\mu^-$
using 1.0\invfb of data~\cite{LHCb-CONF-2012-015}.
The upper limit on the branching fraction was found to be
$\BF(\tau^-\to\mu^+\mu^-\mu^-)<7.8~(6.3) \times 10^{-8}$ at 95\,\%
(90\,\%) C.L, to be compared with the current best
experimental upper limit from the Belle collaboration:
$\BF(\tau^-\to\mu^+\mu^-\mu^-)<2.1 \times 10^{-8}$ at 90\,\% C.L.
As the data sample increases this limit is expected to scale with the
square root of the luminosity, with possible further reduction depending on
improvements in the analysis. The large integrated luminosity that
will be collected by the upgraded LHCb experiment will provide a
sensitivity corresponding to an upper limit of a few times $10^{-9}$~\cite{CERN-LHCC-2011-001}.  

\section{Conclusion}
\label{}

Most scenarios of physics beyond the Standard Model of particle
physics predict measurable effects in the flavor sector, in particular
in rare $B$ meson decays. 
No sign of physics beyond the Standard Model has yet been observed and
stringent limits on its scale have been set.

Sensitive probes for NP are the leptonic decays \Bsdmm and the rare
electroweak penguin decay \BdToKstmm. The most recent measurements
performed by the LHCb collaboration in these two modes are
in good agreement with SM predictions and set strong constraints on
possible extensions of the SM. The isospin asymmetry in the decays
$\decay{B}{\kaon^{(*)} \mup\mun}$ has been measured by several
experiments. These measurements agree with each other and with SM
predictions for the decays with a \Kstarz, but there is a
significant tension with respect to expectations for the decays with a
kaon. Precise calculations are needed to interpret this tension.




\section*{References}
\bibliographystyle{elsarticle-num}
\bibliography{article}   







\end{document}